\documentclass[hyphens,manuscript,screen,nonacm,format=sigconf]{acmart}
\usepackage{soul}
\usepackage{breakurl}

\hypersetup{
           breaklinks=true,   
           colorlinks=true,   
           pdfusetitle=true,  
        }

\acmConference[CI Symposium]{}{September 21--22,
  2023}{Toronto, Canada}
\acmBooktitle{Contextual Integrity Symposium '23, 
  Toronto, Canada}

\title{When PETs misbehave: A Contextual Integrity analysis}
\author{Ero Balsa}
\email{ero.balsa@cornell.edu}
\affiliation{%
  \institution{Cornell Tech}
  \city{New York}
  \state{NY}
  \country{USA}
}
\author{Yan Shvartzshnaider}
\email{yansh@lassonde.yorku.ca}
\affiliation{%
  \institution{York University}
  \city{Toronto}
  \state{ON}
  \country{Canada}
}
\date{June 2023}

\begin{abstract}

Privacy enhancing technologies, or \emph{PETs}, have been hailed as a promising means to protect privacy without compromising on the functionality of digital services.
At the same time, and partly because they may encode a narrow conceptualization of privacy as confidentiality that is popular among policymakers, engineers and the public, PETs risk being co-opted to promote privacy-invasive practices. 

In this paper, we resort to the theory of Contextual Integrity to  explain how privacy technologies may be misused to erode privacy. To illustrate, we consider three PETs and scenarios: anonymous credentials for age verification, client-side scanning for illegal content detection, and homomorphic encryption for machine learning model training. Using the theory of Contextual Integrity, we reason about the notion of privacy that these PETs encode, and show that CI enables us to identify and reason about the limitations of PETs and their misuse, and which may ultimately lead to privacy violations. 
\end{abstract}

\begin{document}

\maketitle

\section{Introduction}

Privacy enhancing technologies, or \emph{PETs}, have been promoted as the key to privacy-by-design, 
as part of the toolbox that any conscientious designer should rely on to devise a privacy-preserving system. 
Much of the allure behind PETs lies on their \emph{positive-sum} quality: rather than imposing a trade-off between functionality and privacy, PETs are hailed as a transformative way to protect privacy without compromising functionality.\footnote{We borrow the concept \emph{positive-sum} from one of the seven foundational principles that Commissioner Cavoukian first identified as key to a privacy-by-design~\cite{cavoukian2009privacy}} 
In many cases, this is largely the case. 
Privacy enhancing technologies, particularly those based on sophisticated cryptographic techniques, enable us, among other functionalities, to communicate online without exposing our messages to the messaging service provider, to run statistic analyses over sensitive data without exposing any of those data and, more recently, to perform COVID-19 exposure notification without the need to track every citizen’s movements and whereabouts~\cite{ermoshina2016end,shokri2015privacy,troncoso2020decentralized}.

At the same time, the excitement for PETs has been co-opted to further an agenda in which PETs are increasingly and disingenuously weaponized to legitimize practices which, at best, are problematic and, at worst, are decidedly privacy-invasive. 
Google’s proposal to do away with third-party cookies and rely on privacy technologies to perform behavioral advertising is a case in point. Revealing little to no information about users’ browsing histories and behavior to third-parties and even to Google itself is certainly more privacy preserving than the current system of tracking everyone everywhere. But is it privacy preserving overall? After all, even with the use of PETs tracking would still happen everywhere, even without cookies, and everyone would still be likely to encounter highly personalized (and potentially manipulative) ads~\cite{cyphers2019google}.

Part of the problem stems from a narrow conceptualization of privacy, reflected both in privacy legislation and popular discourse, where the focus is on protecting personal data. 
Often, the privacy problems that typically garner the most attention and concern relate to the unintended or undesirable disclosure of personal data, with the most potential for immediate and direct harm on individuals~\cite{citron2022privacy,cofone2017privacy}, 
in contrast to more subtle and insidious forms of privacy violations that arise at the population level, as byproducts of the processing of data across many individuals, and which seldom lead to such immediate, noticeable harms~\cite{citron2022privacy,gurses2010pets, viljoen2021relational}.
Hence, even if PETs hide people's identity, personal details and communication from governments and service providers alike, they may still enable surveillance practices that do not necessarily depend on the ability to identify individuals, but rather on the ability to categorize and control populations, e.g.~social sorting~\cite{gurses2010pets,lyon2003surveillance}.

PETs are certainly no panacea and do not, by virtue of their name, automatically protect privacy: it depends on how they are used and the \emph{context} in which they are deployed. 
A poor understanding of PETs' affordances and limitations, 
as well as of key contextual and sociotechnical factors may lead to situations where PETs, rather than protect privacy, may undermine it. 

In this paper, we resort to Nissenbaum's theory of contextual integrity (CI) to perform an analysis of how PETs may lead to or enable privacy violations~\cite{nissenbaum2014respect}. The CI framework prompts us to consider the information flows that PETs either prevent or enable, and to consider whether these flows conform to prevalent privacy norms and expectations. Any information flows that deviate from such norms reveal a prima facie violation of privacy, hence requiring further analysis through the CI heuristic to determine whether they align with contextual values and purposes. 

We analyze three types of PETs and scenarios: anonymous credentials for age verification, client-side scanning for illegal content detection, and homomorphic encryption for machine learning model training. 
We consider the privacy claims often associated to these PETs and compare them with the findings that a CI analysis reveals, thus revealing blind spots and shortcomings that illustrate how PETs may weaken privacy.

More particularly, we show that a) while PETs are designed to \emph{prevent} inappropriate flows, they can also be used to enable (other, equally) inappropriate flows; b) a CI provides a language and framework to reveal and reason about the blind spots and misuses of PETs; and hence, c) to claim that PETs enhance privacy without attending to the particular context in which they are deployed risks enabling inappropriate flows that violate contextual integrity and, as a result, privacy. Moving forward, we hope to further explore how CI may illuminate PETs' design and deployment, to better inform policymakers, privacy engineers, and users.

\section{Background}

G\"{u}rses identifies three \emph{paradigms} within computer science privacy research: privacy as \emph{confidentiality}, privacy as \emph{control} and privacy as \emph{practice}. Within each of these paradigms, researchers rely on a particular conceptualization of privacy and set of assumptions, leading to a distinct set of privacy tools~\cite{gurses2010multilateral}
The privacy as confidentiality paradigm focuses on privacy problems that result from overbearing governments and untrustworthy companies, providing users anonymity and encryption tools as a solution. 
Conversely, the privacy as control paradigm assumes trust in data handlers and institutions, providing users tools to better control how their data is shared and used.
Lastly, within privacy as practice we encounter tools that give users the means to better understand how their data may be used, as well as to negotiate and reconfigure their social boundaries~\cite{gurses2010multilateral,diaz2012understanding}.

Most cryptographic PETs can be inscribed within the privacy as confidentiality paradigm. 
This mirrors the conceptualization of privacy within cryptography, where privacy is a traditional term of art that simply equates to confidentiality, namely, the inability of an adversarial party to obtain a certain piece of information~\cite{boudgoust2023simple,boyle2023must,frederiksen2015privacy}.
Cryptographic PETs thus encode a narrow notion of privacy, namely, privacy as confidentiality, that does not capture the complexity and multifaceted nature of privacy as theorized elsewhere~\cite{mulligan2016privacy,nissenbaum2004privacy}.

As G\"{u}rses argues, \emph{``confidentiality as the way to preserve privacy is a recurring theme in privacy debates generally and in computer science specifically''}~\cite{gurses2010pets}.
Yet she also identifies, drawing from surveillance studies,
the limits of PETs insofar as they enact this narrow notion of privacy as confidentiality~\cite{gurses2010pets}.
PETs can hide people's identity, personal details and communication from governments and service providers while still enabling surveillance practices that do not necessarily depend on the ability to identify individuals, but rather on the ability to categorize and control populations, e.g.~social sorting~\cite{gurses2010pets,lyon2003surveillance}.

This tension between privacy problems that relate to the unintended or undesirable disclosure of personal data, in contrast to privacy problems that arise at the population level has also been acknowledged by privacy scholars, gaining increasing recognition. Privacy scholars have acknowledged the limits of anonymity as an absolute form of privacy protection, as well as the flaws of a individualistic conceptualization of privacy as opposed to a relational notion that understands the social dimension of data~\cite{barocas2014big,nissenbaum2019contextual,viljoen2021relational}.
As G\"{u}rses argues, \emph{``[i]f we accept that data are relational, then we have to reconsider what it means to think within the framework of data protection that only protects “personal information''}~\cite{gurses2010pets}.

\section{PETs through the lens of CI}

In this section, we analyze three instances of PETs deployment that lead to privacy invasions. Using Nissenbaum's framework of Contextual Integrity, we show that a contextual analysis helps us understand and explain why and how, in spite of their name, PETs may be deployed in ways that undermine privacy. 

\subsection{Age verification with anonymous credentials}

Let us consider a privacy-preserving age verification service to prevent minors from accessing certain websites. These services are growing in popularity, largely due to concurrent regulatory efforts in Europe and the US to mandate providers of pornographic content online to perform age-verification~\cite{kayali2023agev,roth2023agev}.

Proposed solutions include verifying credit card details, facial analysis, or using government-issued IDs, such as passports or driver’s licenses. 
These methods are however vulnerable to abuse, e.g.~a pornographic site could be hacked to collect the faces of those visiting the site~\cite{cnil2022agev,kelley2023agev,roth2023agev}.

To mitigate any potential for abuse, and in line with data minimization principles, researchers have proposed the use of solutions that rely on anonymous (or pseudonymous) credentials~\cite{camenisch2019fast,kakvi2023sok,masmoudi2022pima}.
Anonymous credentials are a cryptographic technique that would enable users to certify that they are of age without revealing their date of birth, their precise age, or any other information typically contained in a government-issued ID.
Anonymous credentials further provide guarantees of anonymity and unlinkability, namely, a provider could neither use age-verification alone to identify users, nor track them across multiple sites. 
Hence, this solution would ostensibly be privacy-preserving, as it satisfies the data minimization principle: it only reveals the strictly minimum amount of information age-verification requires~\cite{babel2023bringing,wastlund2012evoking}. 

\subsubsection*{Beyond the technical.}
In spite of providing strong guarantees of anonymity and unlinkability, once embedded in a broader sociotechnical system to support verification, anonymous credentials could be co-opted to weaken privacy overall.
With a verifying infrastructure in place, the market of attribute-verification providers would flourish, creating economic incentives to adopt unnecessary or abusive verification in services that do not require it.

Abusive providers could attempt to legitimize their practices under the cover of using PETs, claiming that such solutions are technically anonymous and unlinkable, and that users can choose the level of data granularity they are comfortable revealing. 
Moreover, nothing would stop a service provider from requesting users to disclose their name, address, age and any other attributes typically requested in sign-up forms today, \emph{in addition to} a request for \emph{anonymous, unlikable} verification of any of those attributes. Much in the same way that Tor users forsake their anonymity if they disclose their name or personal data \emph{while using Tor}, users could be easily duped---if not forced to---reveal then verify personal information. The current landscape of dark patterns for privacy attest to the cunning ability of providers to bypass existing regulation and coerce users into making decisions that go against their best interests~\cite{mathur2019dark}.
Hence, such abusive verification would ultimately weaken the ability of users to forge or manage their own identities online, e.g.~by maintaining several personas or obfuscating details about their true identity.

Applying the CI framework to this scenario can help us reason about privacy in the following way.
Firstly, a CI analysis enables us to flag situations where anonymous credentials may be used to enable inappropriate information flows. For example, we could identify information flows of attributes like users' name, address, or other ID-based credentials to an ``abusive'' recipient service provider, either due to an illegitimate purpose or condition for the information exchange. 

Anonymous credentials are privacy enhancing technologies insofar as they enable certain, \emph{appropriate} flows of information while preventing other, \emph{inappropriate} flows of information, e.g. they reveal that a user is of age (an appropriate flow of information in the context of pornographic content consumption) without revealing their precise date of birth (an unnecessary and therefore inappropriate request). At the same time, their ability to selectively enable and prevent information flows cannot be deemed privacy-enhancing by default, irrespective of context, as their application may disregard key contextual parameters. 

\subsection{Client-side scanning}

The successful deployment of end-to-end encryption in popular instant messaging platforms such as WhatsApp, Telegram or Signal has curtailed the surveillance capabilities of law enforcement agencies: they claim to have become blind to the organization of illegal activity online, as well as the exchange of illegal content~\cite{fbiLawfulAccess,koch2023css,opa2020e2ee}.
As a result, ongoing legislative efforts across several Western countries are pushing towards restrictions on the use of end-to-end encryption~\cite{burgess2022eu,goodin2023signal,pfefferkorn2020earnit}. 

Efforts to weaken encryption have failed in the past, partly thanks to security and privacy experts' warnings about the risks of doing so, as well as fierce opposition from privacy activists and watchdogs~\cite{abelson2021bugs,goodin2023signal,van2013privacy,jarvis2020crypto}.
As a response to this opposition, legislators have turned their attention to client-side scanning~\cite{green2019css,mullin2023earnit,landau2021bugs}.

Client-side scanning (CSS) works by scanning content on a user's device before it is encrypted and transmitted. CSS proposals often rely on some form of specialized hashing, comparing hashes of user content to a database of hashes of illegal or targeted content~\cite{green2019css}. The combination of end-to-end encryption, local computation and content hashing is portrayed as win-win for privacy and security~\cite{noone2022css}. According to this view, it is enough of a privacy guarantee to law-abiding citizens, as no legal content ever leaves their devices, while it enables to detect illegal content and trigger a report to law enforcement. Indeed, CSS does not \emph{technically} break end-to-end encryption: messages are scanned before and/or after they are encrypted and decrypted by the legitimate sender and recipient, respectively. 

\subsubsection*{Beyond the technical.}

Numerous experts have warned of the security and privacy dangers of CSS. Firstly, there is the danger of function creep. Databases of materials to be searched for with CSS could be modified to target content and activities beyond CSAM and terrorism, respectively. Secondly, even assuming such function creep would never take place, experts have identified a myriad of security and privacy issues that malicious parties could exploit in current proposals~\cite{abelson2021bugs}.
Thirdly, and more fundamentally, even if we omit the two points above, CSS violates privacy expectations and contravenes longstanding informational norms. A CI analysis helps us illustrate that even if CSS does not {\em technically} break end-to-end encryption and relies on supposedly privacy-preserving techniques such as hashing and local, distributed computation, it still (prima facie) violates privacy. 

At first glance, by virtue of relying on local scanning and implementing end-to-end encryption, CSS prevents the mass flow of information from every citizen to law enforcement. In that sense, it would seem that CSS is in line with contextual integrity, as it prevents the mass, indiscriminate flow of information from people to government agencies. 
However, upon closer examination, and in apparent sleight of hand, CSS still enables a mass flow of information from every citizen to law enforcement, in addition to the explicit, targeted flows of information from allegedly law-infringing citizens.

Detection of illegal content triggers an explicit notification to law enforcement. At the same time, non-detection implicitly signals that, if CSS is working as expected, no illegal content has been detected on a user's device, thus revealing valuable information to the government, even in the absence of an explicit notification.
It is thus no surprise that top security and privacy researchers have described CSS as \emph{``bulk intercept, albeit automated and distributed''}~\cite{abelson2021bugs}.

Applying CI to analyze these flows further reveals how they constitute a prima facie violation of privacy.
In liberal societies, to perform wiretapping legally, law enforcement agencies need to present sufficient evidence in front of a judge, who decides, upon examining the merits of the evidence, whether or not to issue a wiretap warrant. Hence, in this case, \emph{``with a warrant''} represents the transmission principle that enables the flow of information from a suspect's device to law enforcement. 
Conversely, with CSS no such evidence or warrant are necessary, contravening the transmission principle \emph{``with a warrant''} that preserves contextual integrity. With CSS information automatically flows from \emph{all} devices: explicitly where illegal content is found, implicitly where it is not. Hence, a prima facie violation of contextual integrity takes place, as the new transmission principle tramples longstanding informational norms.

\subsection{Privacy-preserving machine learning}

The development of ML models has raised numerous privacy concerns, such as the lack of consent in scrapping publicly available data on the web to train ML models, as well as the potential for ML models to reveal their training data because of their memorization properties~\cite{bartlett2021beyond,de2021critical,kerry2020ai,papernot2018sok}.
To address these problems, researchers have proposed a number of PETs, often combining both cryptographic and non-cryptographic tools, such as federated learning, multiparty computation (MPC), homomorphic encryption (HE) or differential privacy (DP), among others~\cite{mohassel2017secureml,sebert2022protecting,xu2021privacy}.
To illustrate how CI can help us illuminate the privacy concerns left unaddressed by these PETs, as well as the more insidious privacy violations they may enable, let us consider the use of HE in the development of large language models (LLMs). 

HE is a cryptographic technique that enables operating on encrypted data without having to decrypt the data first. This property makes HE a promising security and privacy technology, as clients may outsource computations over sensitive or confidential data to third parties, who can operate over those encrypted data without having access to the underlying (\emph{plaintext}) data. 

Let us consider an ML provider that wishes to train an LLM over sensitive data, e.g.~a hospital's patient data. 
With HE, the provider could collect encrypted data encrypted and use it to train an LLM. The training data would remain encrypted throughout the training process. Ostensibly, this solution would address any privacy concerns, as the provider would obtain the LLM without ever having access to the underlying training data~\cite{castro2023chatgpt}.

\subsubsection*{Beyond the technical}
HE provides certain privacy guarantees, insofar as it prevents  the ML provider from having access to the underlying data, thus preventing data breaches and other types of abuse. 
However, assuming that HE automatically solves all of LLMs' privacy issues relies on contextual assumptions that HE proponents seldom make explicit. 

A CI-based analysis forces us to reckon with those contextual assumptions, namely, the contexts in which the information that such an LLM requires and generates flows. Hence, we must consider the ends and values that the flow of information from individuals supplying training data to the provider who builds the model promotes, as well as the information flows that the model may subsequently engender.
CI prompts us to consider the \emph{actors} involved in the training process, namely, who builds the model with whose data, who uses the LLM and to whom the information from the LLM flows. 

Hence, we could deem that a hospital hoping to improve patient care by simulating interactions with patients during hospital staff training would have a legitimate claim to an LLM built on clinical data. 
Conversely, an advertiser hoping to improve its prediction models to determine when to target consumers with healthcare services or drugs would not. 

CI also enables us to identify the illegitimate redeployment or repurposing of existing, legitimate models as a privacy violation, such as in the event the hospital sells to an advertiser the LLM it legitimately built to improve personnel training.
In this particular case, the redeployment reveals a prima facie privacy violation, as data that patients provided to the hospital in the context of improving health care is repurposed in a commercial context, with no obvious contribution to the ends and values of the original context.

\subsection{CI's conservative bias}
\label{ciConservative}

In our analysis of the three use cases above, we have shown how  the deployment of PETs leads to \emph{prima facie} privacy violations, i.e.~we have observed how PETs enable flows of information that contravene established information norms, ostensibly leading to a privacy violation. 
It is however important to note that whether these information flows ultimately represent a privacy violation depends on whether they contribute to or promote the ends and values of the context in which they take place, among other considerations.

As Nissenbaum acknowledges, the CI framework has a bias towards established norms~\cite{nissenbaum2004privacy}. We use CI to determine legitimacy and appropriateness of information flows by subjecting them to existing privacy norms. However, in cases where entrenched informational norms are in fact undermining the ends and values of the particular context they govern, a breach would ultimately be beneficial to society. The CI heuristic helps navigate these cases~\cite{nissenbaum2014respect}. After identifying a norm-breaching flow we can evaluate its appropriateness through three levels of analysis:  a) assess the implication of novel flows on the interests of key stakeholders (costs and benefits) b) evaluate of respective values (moral, social, and political), c) analyze the contribution of the flow to the context values, ends, and purposes.

\section{Takeaways}

In this paper we have subjected three privacy technologies to a CI analysis to reveal how, in apparent contradiction to their primary purpose, they may lead to privacy violations. 
Our analysis of anonymous credentials shows that, if used where no credentials are needed---anonymous or otherwise---they enable inappropriate flows of information that may violate privacy. 
Our analysis of client-side scanning illustrates how the joint deployment of end-to-end encryption, local computation and hashing may not directly expose people's messages to law enforcement, but still enables a mass flow of information about their communications that violates informational norms. 
Lastly, our analysis of homomorphic encryption to protect the data used to train ML models highlights the limits of confidentiality, insofar as the use of such ML models may undermine the ends and values of the context in which the training data were generated. 

Indeed, all three cases illustrate how the protection of \emph{some} data or, in CI parlance, the preemption of some inappropriate flows of information, can be used as a cover to enable potentially privacy-invasive functionalities, i.e.~enabling the flow of other flows of information that contravene longstanding informational norms. 
Hence, to proclaim that privacy enhancing technologies enhance privacy without attending to the particular context in which they are deployed risks enabling inappropriate flows that violate contextual integrity and, as a result, privacy. 
Instead, we must consider the flow of information between specific actors (sender, subject, recipient) under specific conditions (transmission principle) to determine whether such deployments may promote or undermine the ends and values of the context in which they take place, and therefore, privacy.

Our CI analysis also adds weight to the raising tide of voices that warn against a myopic conceptualization of privacy as confidentiality, or too narrowly focused on the protection of personal or sensitive data, disregarding the social nature of privacy and the relationality of data. 

Ultimately, this paper illustrates how, by prompting us to be more attentive to the context in which PETs operate, a CI analysis can help us reason about the blind spots and limitations of these technologies, hopefully leading to more principled and constructive PET design and deployment.

\section*{}
\subsubsection*{Acknowledgements}

We would like to thank the organizers and participants of CNIL's Privacy Research Day 2023, in particular Vincent Toubiana and Mehdi Arfaoui, for discussions related to the use and misuse of privacy technologies. The use cases in this paper are taken directly from work that Jain, Laurent and S\'{e}bert presented at the event~\cite{jain2022adversarial,masmoudi2022pima,sebert2022protecting}.

\bibliography{misbehavingPETs}

\begin{thebibliography}{10}

\bibitem{abelson2021bugs}
Hal Abelson, Ross Anderson, Steven~M Bellovin, Josh Benaloh, Matt Blaze, Jon
  Callas, Whitfield Diffie, Susan Landau, Peter~G Neumann, Ronald~L Rivest,
  et~al.
\newblock Bugs in our pockets: The risks of client-side scanning.
\newblock {\em arXiv preprint arXiv:2110.07450}, 2021.

\bibitem{babel2023bringing}
Matthias Babel and Johannes Sedlmeir.
\newblock Bringing data minimization to digital wallets at scale with
  general-purpose zero-knowledge proofs.
\newblock {\em arXiv preprint arXiv:2301.00823}, 2023.

\bibitem{barocas2014big}
Solon Barocas and Helen Nissenbaum.
\newblock Big data’s end run around anonymity and consent.
\newblock {\em Privacy, big data, and the public good: Frameworks for
  engagement}, 1:44--75, 2014.

\bibitem{bartlett2021beyond}
Matt Bartlett.
\newblock Beyond privacy: Protecting data interests in the age of artificial
  intelligence.
\newblock {\em Law, Technology and Humans}, 3(1):96--108, 2021.

\bibitem{boudgoust2023simple}
Katharina Boudgoust and Peter Scholl.
\newblock Simple threshold (fully homomorphic) encryption from lwe with
  polynomial modulus.
\newblock {\em Cryptology ePrint Archive}, 2023.

\bibitem{boyle2023must}
Elette Boyle, Ran Cohen, Deepesh Data, and Pavel Hub{\'a}{\v{c}}ek.
\newblock Must the communication graph of {MPC} protocols be an expander?
\newblock {\em Journal of Cryptology}, 36(3):20, 2023.

\bibitem{burgess2022eu}
Matt Burgess.
\newblock {EU} plan to scan private messages for child abuse images puts
  encryption at risk.
\newblock Wired. Online at
  \url{https://www.wired.com/story/europe-csam-scanning-law-chat-encryption/},
  May 2022.

\bibitem{camenisch2019fast}
Jan Camenisch, Manu Drijvers, Petr Dzurenda, and Jan Hajny.
\newblock Fast keyed-verification anonymous credentials on standard smart
  cards.
\newblock In {\em ICT Systems Security and Privacy Protection: 34th IFIP TC 11
  International Conference, SEC 2019, Lisbon, Portugal, June 25-27, 2019,
  Proceedings 34}, pages 286--298. Springer, 2019.

\bibitem{castro2023chatgpt}
Chiara Castro.
\newblock This company believes to have the solution to {ChatGPT} privacy
  problems.

\bibitem{cavoukian2009privacy}
Ann Cavoukian.
\newblock Privacy by design.
\newblock 2009.

\bibitem{citron2022privacy}
Danielle~Keats Citron and Daniel~J. Solove.
\newblock Privacy harms.
\newblock {\em BUL Rev.}, 102:793, 2022.

\bibitem{cnil2022agev}
CNIL.
\newblock Online age verification: balancing privacy and the protection of
  minors.
\newblock Online at
  \url{https://www.cnil.fr/en/online-age-verification-balancing-privacy-and-protection-minors},
  September 2022.

\bibitem{cofone2017privacy}
Ignacio~N Cofone and Adriana~Z Robertson.
\newblock Privacy harms.
\newblock {\em Hastings LJ}, 69:1039, 2017.

\bibitem{cyphers2019google}
Bennett Cyphers.
\newblock Don't play in {G}oogle's {P}rivacy {S}andbox.
\newblock EFF. Online at
  \url{https://www.eff.org/deeplinks/2019/08/dont-play-googles-privacy-sandbox-1},
  August 2019.

\bibitem{de2021critical}
Emiliano De~Cristofaro.
\newblock A critical overview of privacy in machine learning.
\newblock {\em IEEE Security \& Privacy}, 19(4):19--27, 2021.

\bibitem{diaz2012understanding}
Claudia Diaz and Seda G{\"u}rses.
\newblock Understanding the landscape of privacy technologies.
\newblock {\em Proceedings of the information security summit}, 12:58--63,
  2012.

\bibitem{ermoshina2016end}
Ksenia Ermoshina, Francesca Musiani, and Harry Halpin.
\newblock End-to-end encrypted messaging protocols: An overview.
\newblock In {\em Internet Science: Third International Conference, INSCI 2016,
  Florence, Italy, September 12-14, 2016, Proceedings 3}, pages 244--254.
  Springer, 2016.

\bibitem{fbiLawfulAccess}
{Federal Bureau of Investigation}.
\newblock The lawful access challenge.
\newblock Online at \url{https://www.fbi.gov/about/mission/lawful-access}.

\bibitem{frederiksen2015privacy}
Tore~Kasper Frederiksen, Jesper~Buus Nielsen, and Claudio Orlandi.
\newblock Privacy-free garbled circuits with applications to efficient
  zero-knowledge.
\newblock In {\em Advances in Cryptology-EUROCRYPT 2015: 34th Annual
  International Conference on the Theory and Applications of Cryptographic
  Techniques, Sofia, Bulgaria, April 26-30, 2015, Proceedings, Part II}, pages
  191--219. Springer, 2015.

\bibitem{goodin2023signal}
Dan Goodin.
\newblock Signal {CEO}: We ``1,000\% won't participate'' in {UK} law to weaken
  encryption.
\newblock Ars Technica. Online at
  \url{https://arstechnica.com/information-technology/2023/02/signal-vows-to-defy-uk-legislation-that-puts-e2e-encryption-in-the-crosshairs/},
  February 2023.

\bibitem{green2019css}
Matthew Green.
\newblock Can end-to-end encrypted systems detect child sexual abuse imagery?
\newblock Online at
  \url{https://blog.cryptographyengineering.com/2019/12/08/on-client-side-media-scanning/},
  December 2019.

\bibitem{gurses2010multilateral}
Seda G{\"u}rses.
\newblock {\em Multilateral privacy requirements analysis in online social
  network services}.
\newblock PhD thesis, KU Leuven, 2010.

\bibitem{gurses2010pets}
Seda G{\"u}rses.
\newblock {PETs} and their users: a critical review of the potentials and
  limitations of the privacy as confidentiality paradigm.
\newblock {\em Identity in the Information Society}, 3:539--563, 2010.

\bibitem{jain2022adversarial}
Shubham Jain, Ana-Maria Crețu, and Yves-Alexandre de~Montjoye.
\newblock Adversarial detection avoidance attacks: Evaluating the robustness of
  perceptual hashing-based client-side scanning.
\newblock In {\em 31st USENIX Security Symposium (USENIX Security 22)}, pages
  2317--2334, 2022.

\bibitem{jarvis2020crypto}
Craig Jarvis.
\newblock {\em Crypto {W}ars: {T}he Fight for Privacy in the Digital Age: a
  Political History of Digital Encryption}.
\newblock CRC Press, 2020.

\bibitem{kakvi2023sok}
Saqib~A Kakvi, Keith~M Martin, Colin Putman, and Elizabeth~A Quaglia.
\newblock Sok: Anonymous credentials.
\newblock In {\em International Conference on Research in Security
  Standardisation}, pages 129--151. Springer, 2023.

\bibitem{kayali2023agev}
Laura Kayali.
\newblock No porn, no instagram for kids: France doubles down on age
  verification.
\newblock Politico. Online at
  \url{https://www.politico.eu/article/no-porn-no-instagram-for-kids-france-doubles-down-age-verification-emmanuel-macrons-nick-clegg/},
  February 2023.

\bibitem{kelley2023agev}
JASON KELLEY and ADAM SCHWARTZ.
\newblock Age verification mandates would undermine anonymity online.
\newblock EFF. Online at
  \url{https://www.eff.org/deeplinks/2023/03/age-verification-mandates-would-undermine-anonymity-online},
  March 2023.

\bibitem{kerry2020ai}
Cameron~F. Kerry.
\newblock Protecting privacy in an {AI}-driven world.
\newblock The Brookings Institution. Online at
  \url{https://www.brookings.edu/research/protecting-privacy-in-an-ai-driven-world/},
  February 2020.

\bibitem{koch2023css}
Richie Koch.
\newblock Why client-side scanning isn’t the answer.
\newblock Online at
  \url{https://proton.me/blog/why-client-side-scanning-isnt-the-answer},
  January 2023.

\bibitem{landau2021bugs}
Susan Landau.
\newblock Bugs in our pockets: The risks of client-side scanning.
\newblock Lawfare. Online at
  \url{https://www.lawfareblog.com/bugs-our-pockets-risks-client-side-scanning},
  October 2021.

\bibitem{lyon2003surveillance}
David Lyon.
\newblock {\em Surveillance as social sorting: Privacy, risk, and digital
  discrimination}.
\newblock Psychology Press, 2003.

\bibitem{masmoudi2022pima}
Souha Masmoudi, Maryline Laurent, and Nesrine Kaaniche.
\newblock {PIMA}: {A} {P}rivacy-preserving {I}dentity management system based
  on an unlinkable {MA}lleable signature.
\newblock {\em Journal of Network and Computer Applications}, 208:103517, 2022.

\bibitem{mathur2019dark}
Arunesh Mathur, Gunes Acar, Michael~J Friedman, Eli Lucherini, Jonathan Mayer,
  Marshini Chetty, and Arvind Narayanan.
\newblock Dark patterns at scale: Findings from a crawl of 11k shopping
  websites.
\newblock {\em Proceedings of the ACM on Human-Computer Interaction},
  3(CSCW):1--32, 2019.

\bibitem{mohassel2017secureml}
Payman Mohassel and Yupeng Zhang.
\newblock Secure{ML}: A system for scalable privacy-preserving machine
  learning.
\newblock In {\em 2017 IEEE symposium on security and privacy (SP)}, pages
  19--38. IEEE, 2017.

\bibitem{mulligan2016privacy}
Deirdre~K Mulligan, Colin Koopman, and Nick Doty.
\newblock Privacy is an essentially contested concept: a multi-dimensional
  analytic for mapping privacy.
\newblock {\em Philosophical Transactions of the Royal Society A: Mathematical,
  Physical and Engineering Sciences}, 374(2083):20160118, 2016.

\bibitem{mullin2023earnit}
Joe Mullin.
\newblock The {EARN IT} bill is back, seeking to scan our messages and photos.
\newblock {EFF}. Online at
  \url{https://www.eff.org/deeplinks/2023/04/earn-it-bill-back-again-seeking-scan-our-messages-and-photos},
  April 2023.

\bibitem{nissenbaum2004privacy}
Helen Nissenbaum.
\newblock Privacy as contextual integrity.
\newblock {\em Wash. L. Rev.}, 79:119, 2004.

\bibitem{nissenbaum2014respect}
Helen Nissenbaum.
\newblock Respect for context as a benchmark for privacy online: What it is and
  isn’t.
\newblock {\em Cahier de prospective}, 19, 2014.

\bibitem{nissenbaum2019contextual}
Helen Nissenbaum.
\newblock Contextual integrity up and down the data food chain.
\newblock {\em Theoretical Inquiries in Law}, 20(1):221--256, 2019.

\bibitem{noone2022css}
Greg Noone.
\newblock Is client-side scanning the future of content moderation?
\newblock TechMonitor. Online at
  \url{https://techmonitor.ai/policy/privacy-and-data-protection/client-side-scanning-content-moderation},
  August 2022.

\bibitem{papernot2018sok}
Nicolas Papernot, Patrick McDaniel, Arunesh Sinha, and Michael~P Wellman.
\newblock Sok: Security and privacy in machine learning.
\newblock In {\em 2018 IEEE European Symposium on Security and Privacy
  (EuroS\&P)}, pages 399--414. IEEE, 2018.

\bibitem{pfefferkorn2020earnit}
Riana Pfefferkorn.
\newblock The {EARN IT} act: How to ban end-to-end encryption without actually
  banning it.
\newblock Online at
  \url{https://cyberlaw.stanford.edu/blog/2020/01/earn-it-act-how-ban-end-end-encryption-without-actually-banning-it},
  January 2020.

\bibitem{roth2023agev}
Emma Roth.
\newblock Online age verification is coming, and privacy is on the chopping
  block.
\newblock The Verge. Online at
  \url{https://www.theverge.com/23721306/online-age-verification-privacy-laws-child-safety},
  May 2023.

\bibitem{sebert2022protecting}
Arnaud~Grivet S{\'e}bert, Renaud Sirdey, Oana Stan, and C{\'e}dric
  Gouy-Pailler.
\newblock Protecting data from all parties: Combining {FHE} and {DP} in
  federated learning.
\newblock {\em arXiv preprint arXiv:2205.04330}, 2022.

\bibitem{shokri2015privacy}
Reza Shokri and Vitaly Shmatikov.
\newblock Privacy-preserving deep learning.
\newblock In {\em Proceedings of the 22nd ACM SIGSAC conference on computer and
  communications security}, pages 1310--1321, 2015.

\bibitem{troncoso2020decentralized}
Carmela Troncoso, Mathias Payer, Jean-Pierre Hubaux, Marcel Salath{\'e}, James
  Larus, Edouard Bugnion, Wouter Lueks, Theresa Stadler, Apostolos Pyrgelis,
  Daniele Antonioli, et~al.
\newblock Decentralized privacy-preserving proximity tracing.
\newblock {\em arXiv preprint arXiv:2005.12273}, 2020.

\bibitem{opa2020e2ee}
{U.S. Department of Justice. Office of Public Affairs}.
\newblock International statement: End-to-end encryption and public safety.
\newblock Online at
  \url{https://www.justice.gov/opa/pr/international-statement-end-end-encryption-and-public-safety},
  October 2020.

\bibitem{van2013privacy}
Joris~VJ Van~Hoboken and Ira~S. Rubinstein.
\newblock Privacy and security in the cloud: {S}ome realism about technical
  solutions to transnational surveillance in the post-{S}nowden era.
\newblock {\em Me. L. Rev.}, 66:487, 2013.

\bibitem{viljoen2021relational}
Salome Viljoen.
\newblock A relational theory of data governance.
\newblock {\em Yale LJ}, 131:573, 2021.

\bibitem{wastlund2012evoking}
Erik W{\"a}stlund, Julio Angulo, and Simone Fischer-H{\"u}bner.
\newblock Evoking comprehensive mental models of anonymous credentials.
\newblock In {\em Open Problems in Network Security: IFIP WG 11.4 International
  Workshop, iNetSec 2011, Lucerne, Switzerland, June 9, 2011, Revised Selected
  Papers}, pages 1--14. Springer, 2012.

\bibitem{xu2021privacy}
Runhua Xu, Nathalie Baracaldo, and James Joshi.
\newblock Privacy-preserving machine learning: Methods, challenges and
  directions.
\newblock {\em arXiv preprint arXiv:2108.04417}, 2021.

\end{thebibliography}
\bibliographystyle{plain}

\end{document}